\documentclass{iopart}

\usepackage{graphicx}% Include figure files
\usepackage{dcolumn}% Align table columns on decimal point
\usepackage{bm}% bold math
\usepackage{iopams}
\usepackage{multirow}

\begin{document}

\newcommand{\ground}{$5{\rm s}^{2}\,^1{\rm S}_0$~}
\newcommand{\intermed}{$5{\rm s}5{\rm p}\,^1{\rm P}_1$~}
\newcommand{\probe}{$5{\rm s}^{2}\,^1{\rm S}_0 \rightarrow 5{\rm s}5{\rm p}\,^1{\rm P}_1$~}

\title{Modulation-free pump-probe spectroscopy of strontium atoms}

\author{C Javaux, I G Hughes, G Lochead, J Millen and  M P A Jones}

\address{Department of Physics, Durham University, Rochester Building, South Road, Durham DH1 3LE, United Kingdom}\ead{m.p.a.jones@durham.ac.uk}

\submitto{\JPB}
\begin{abstract} We have performed polarization spectroscopy and sub-Doppler DAVLL on the \probe transition of atomic strontium. Both techniques generated a dispersion-type lineshape suitable for laser stabilization, without the need for frequency modulation. In both cases the signal is generated primarily by saturation effects, rather than optical pumping. The dependence of the amplitude and gradient on  intensity and magnetic field were also investigated. 
\end{abstract}
\maketitle

\section{Introduction}

The stabilization, or ``locking'', of a laser frequency to an atomic transition is an essential technique for experiments in laser cooling and precision metrology. The starting point for laser locking is a spectroscopic technique which produces a lineshape with a zero crossing at the atomic resonance, providing an ``error signal'' that can be used to correct for frequency deviations. Suitable lineshapes with sub-Doppler spectroscopic resolution can be produced by using frequency modulated laser light, as in FM spectroscopy \cite{bjorklund80} or modulation transfer spectroscopy \cite{shirley82,mccarron07}. However it is not always straightforward, or desirable, to use frequency modulated light.  If the modulation is generated by modulating the injection current of a diode laser, then it is present on all the laser beams used in the experiments, which can cause problems for some experiments. For other laser sources, or where residual frequency modulation cannot be tolerated, a costly external modulator must be used. To get around this, spectroscopic techniques that generate a suitable lineshape directly, without the need for frequency modulation, have been developed.

In this paper we apply two of these techniques, polarization spectroscopy \cite{wieman76,demtroder} and sub-Doppler DAVLL (Dichroic Atomic Vapour Laser Locking) \cite{wasik02},  
to the  alkaline earth metal strontium. There is increasing interest in the spectroscopy and laser cooling of alkaline-earth elements such as strontium, motivated by applications in precision frequency metrology \cite{blatt08} and ultra-cold Rydberg and plasma physics \cite{simien04}.  We perform spectroscopy on the \probe cooling transition at 461\,nm, where frequency-doubled laser sources must be used and FM methods are difficult to implement. In previous work with strontium, polarization spectroscopy was used to observe the narrow intercombination line at 689\,nm \cite{tino92}. Both techniques have been extensively characterized for the alkali metals Rb and Cs \cite{pearman02,harris06,harris07}. We find that the electronic structure of strontium leads to important differences. Notably, for the bosonic isotopes, the \ground state has no degeneracy. The polarization spectroscopy signal is due only to the anisotropic saturation of the transition by the pump beam. This is  in contrast to the alkali metals, where optical pumping is the dominant mechanism for generating the circular birefringence. We show that this technique produces a dispersion profile suitable for laser locking, and compare this with the signal produced by the  sub-Doppler DAVLL technique. The  stability of the zero crossing is evaluated by comparison with saturated absorption spectroscopy.

\section{Polarization spectroscopy}

\begin{figure}
\centering
\includegraphics{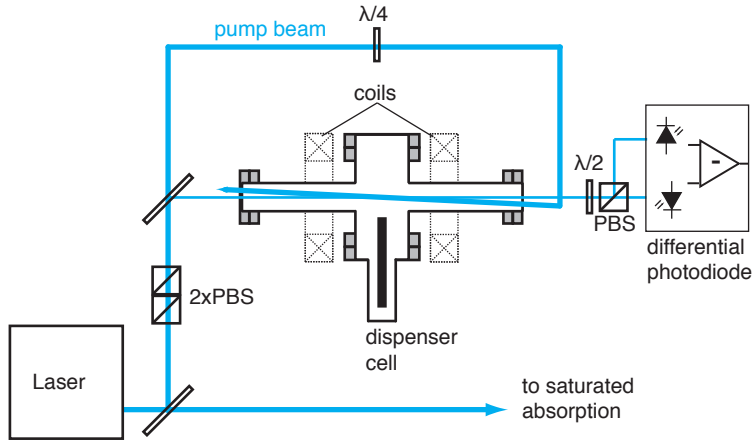}
\caption{\label{setup}(Colour online) Experimental layout for polarization spectroscopy. Polarizing beam splitter cubes (PBS) are used to set and analyse the polarization. For sub-Doppler DAVLL, the $\lambda/2$ and $\lambda/4$ plates are exchanged and a magnetic field is applied along the axis using a pair of coils (shown dotted).}
\end{figure}

In polarization spectroscopy a circularly polarized pump beam induces a birefringence in the medium, which is interrogated using a counterpropagating, linearly polarized probe beam at the same frequency. The experimental arrangement is shown in figure \ref{setup}. The output signal is the intensity difference between the two outputs of the polarizing beamsplitter cube. The probe beam is initially linearly polarized at $\phi=\pi/4$ to the axes of the cube such that the difference signal is zero in the absence of the pump beam. In the presence of the pump beam, the birefringence induced in the medium causes the probe beam polarization to be rotated  through a small angle $\Phi$, giving rise to a non-zero output signal. Expressions for the polarization spectroscopy signal in this configuration were obtained in \cite{pearman02}. Neglecting the birefringence of the cell windows \footnote{Birefringence in the windows leads to a small additional  rotation that can be compensated by adjusting slightly away from $\phi=\pi/4$. Variations can cause a slow drift of the laser lock point.}, the polarization spectroscopy signal after passing through a length $L$ of the medium is
\begin{equation} \label{angle}
I_{\mathrm{signal}} = I_0 e^{-\alpha L} \cos \left( 2\phi + 2\Phi(x) \right)\,,
\end{equation}
with the rotation angle  $\Phi(x)$ given by
\begin{equation}\Phi(x) =  L \frac{\Delta \alpha_0}{2}  \frac{x}{1+x^2} \,,
\end{equation}
where $\Delta \alpha_0$ is the on-resonance difference in absorption coefficient  between the left- and right-hand circularly polarized components of the probe beam, and $x=2(w-w_0)/\Gamma^\prime$ is the detuning in units of half the power-broadened linewidth $\Gamma^\prime$  \footnote{The natural linewidth of the \probe transition is  $\Gamma = 2 \pi \times 32 \,$MHz.}. 

The important parameter that sets the size of the polarization spectroscopy signal is the difference in line centre absorption $\Delta \alpha_0$ induced by the pump beam. In the alkali metals, this difference in absorption arises from two distinct mechanisms: saturation and optical pumping. At high intensity ($I\gg I_{\mathrm{sat}}$) , the circularly polarized pump beam saturates the $\sigma_+$  (or $\sigma_-$) transition, leading to reduced absorption of the corresponding component of the probe beam. At low intensity ($I\lesssim I_{\mathrm{sat}}$) , optical pumping can redistribute the ground state population among the magnetic sublevels, which then do not couple equally to the two circularly polarized components of the probe. With increasing intensity, the dispersion signal can actually change sign as the saturation mechanism dominates over optical pumping \cite{pearman02,harris06}. In the alkaline earths, the situation is different. The even isotopes have no nuclear spin, and the \probe transition is a simple $J=0 \rightarrow J=1$ transition. In this case the polarization spectroscopy signal arises uniquely from saturation as there is no optical pumping. For the odd isotopes, hyperfine structure and magnetic sublevels play a role, and optical pumping can take place.
\begin{figure}
\centering
\includegraphics{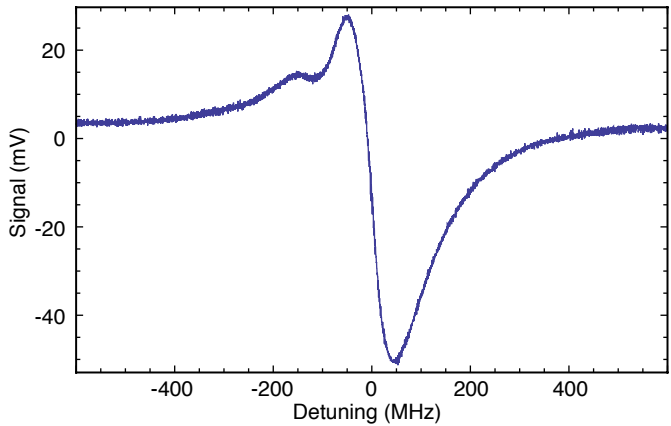}
\caption{\label{polgraph}The polarization spectroscopy signal, obtained with a pump beam intensity of 0.93 W\,cm$^{-2}$.}
\end{figure}

Our experiment was set up as shown in figure \ref{setup}. The 461~nm laser light was produced by a commercial frequency doubled diode laser system (Toptica-SHG). A novel  dispenser-based vapour cell \cite{bridge09}  was used for the measurements. A broad, weakly collimated jet of strontium atoms is produced by electrically heating a commercial vapour source. The pump and probe beams had $1/e^2$ radii of $0.67\pm0.05$\,mm and $0.39\pm0.05$\,mm in the horizontal and vertical directions respectively and were overlapped inside the cell with a small crossing angle ($\sim 20$~mrad).  The probe beam power was fixed at 41~$\mathrm{\mu W}$. An amplified differential photodiode was used to detect the light in the output arms of the PBS. A pair of coils placed symmetrically around the cell allowed a magnetic field to be applied along the direction of the laser beams. We also performed simultaneous saturated absorption spectroscopy using a second dispenser cell. The frequency axis of the laser scan was calibrated by fitting these saturated absorption spectra and using the known natural abundances, isotope shifts and hyperfine splittings \cite{mauger08}. 
\begin{figure}
\centering
\includegraphics{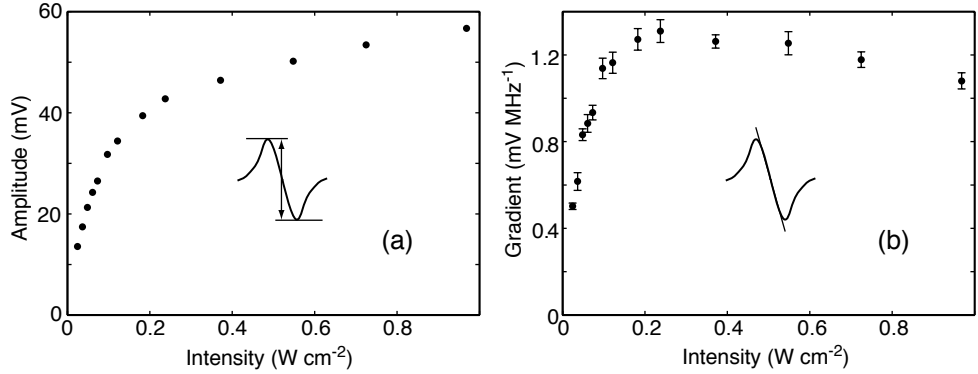}
\caption{\label{ampgrad}Variation of the polarization spectroscopy signal amplitude (a) and gradient (b) with the pump beam intensity. Each point shows the mean value and standard error of five successive measurements. In (a) the size of this error bar is indicated by the size of the data points, and a 1.7$\%$ statistical error on the frequency axis calibration is included in the error bars for the gradient in (b). For the \probe transition, the theoretical saturation intensity is 0.043 W~cm$^{-2}$.}
\end{figure}
An example of the polarization spectroscopy signal that we obtained is shown in figure \ref{polgraph}. A sharp dispersion shaped feature associated with the dominant (83\,$\%$ abundant) $^{88}\mathrm{Sr}$ isotope is clearly visible. This strong feature is ideal for laser locking. On the negative detuning side, contributions from the other isotopes are also visible. The peak signal is also lower on this side of resonance. The shape of the signal here is very dependent on the pump and probe polarization, and is probably due to optical  hyperfine pumping effects in the $^{87}$Sr isotope. The zero-crossing is displaced from the centre of the $^{88}$Sr resonance by approximately 10~MHz. This offset is important for laser locking, and it can be varied (and brought to zero) by making slight adjustments to the angle $\phi$ of the $\lambda/2$ plate. We have investigated the stability of this offset, and we find that the RMS variation of the offset during a $\sim 1$ hour period is 0.8\,MHz. The variation of the amplitude and gradient of the polarization spectroscopy signal is shown in figure \ref{ampgrad}. At high pump intensity ($I\gg I_{\mathrm{sat}}$), the amplitude saturates as expected, while the gradient reaches a maximum before decreasing again as power broadening becomes important.

To make a quantitative model of the polarization spectroscopy signal we must take into account the spatial profile of the beam and the time-dependence of the anisotropy. For rubidium vapour in thermal equilibrium, the time-dependent anisotropy has recently been calculated analytically and numerically \cite{do08}. In our experiment, the velocity distribution of atoms in the jet from the dispenser source is not well characterized, and so a quantitative simulation is beyond the scope of this work. However, we can use  (\ref{angle}) to estimate the rotation angle. From the peak signal at positive detuning, we estimate the maximum rotation angle $\Phi$  to be $\sim2$~mrad.

\section{Sub-Doppler DAVLL}
\begin{figure}
\centering
\includegraphics{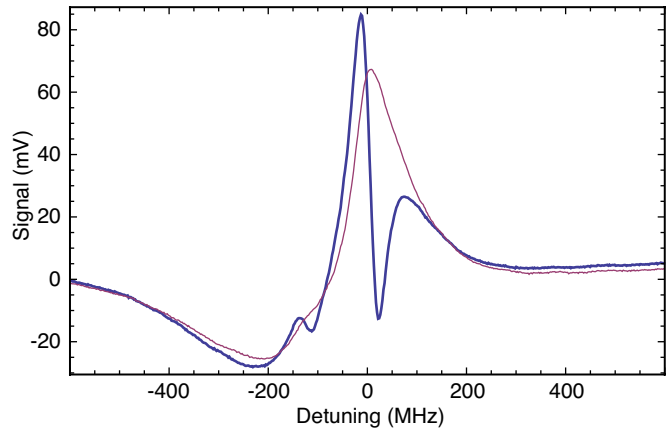}
\caption{\label{subdop}(Colour online) An example sub-Doppler DAVLL signal (thick blue line), obtained with a pump beam intensity of 0.42~W\,cm$^{-2}$ and a magnetic field of 8.5~G. A sharp sub-Doppler dispersion feature is superimposed on the Doppler-broadened background DAVLL signal (thin red line).}
\end{figure}

In order to obtain sub-Doppler DAVLL signals, the half-wave plate before the analyzer cube is replaced with a quarter-wave plate. The axes of this waveplate are adjusted to be at $\pi/4$ with respect to the initial linear polarization. In this configuration, the two circular components of the linearly polarized probe beam exit from different ports of the analyzer cube. Therefore the difference in intensity between the two outputs corresponds to the difference in absorption on the $\sigma^+$ and $\sigma^-$ transitions (circular dichroism). A magnetic field applied along the axis of the probe beam lifts the degeneracy between the $m_J = 0,\pm1$ sublevels of the \intermed excited state, Zeeman shifting the centres of the $\sigma^+$ and $\sigma^-$ transitions equally but in opposite directions.  Hence above or below the zero-field resonance one circular component is absorbed more than the other, while exactly on resonance the difference signal is zero. Conventional DAVLL \cite{sorel94,corwin98} uses a probe beam only, and it is the Doppler-broadened absorption line that must be Zeeman split by the magnetic field.  In sub-Doppler DAVLL, a linearly  polarized pump beam is added (by removing the quarter-wave plate in figure \ref{setup}), and sharp dispersion features are obtained at much lower magnetic fields by splitting the sub-Doppler spectral lines. In strontium, these sub-Doppler features originate solely from saturation, whereas in the alkali metals hyperfine pumping is the dominant mechanism \cite{smith04}.
\begin{figure}
\centering
\includegraphics{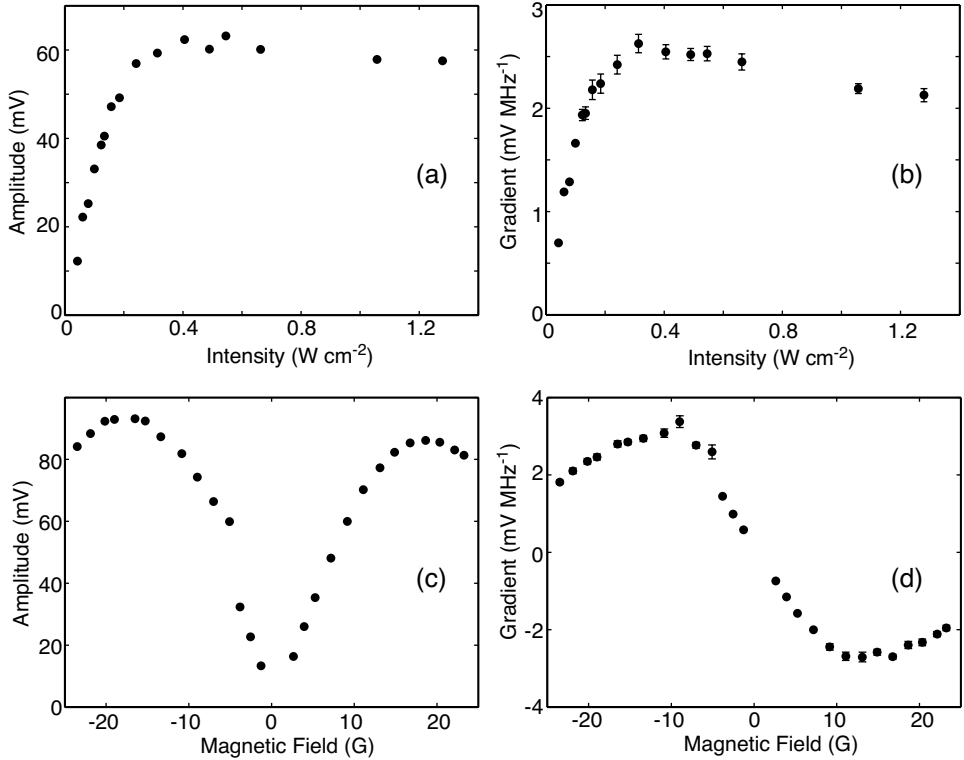}
\caption{\label{BandP}(a and b)Variation of the amplitude (a) and gradient (b) of the sub-Doppler DAVLL signal as a function of intensity, with the magnetic field fixed at  9.5~G. (c and d) Variation of the amplitude (c) and gradient (d) of the sub-Doppler DAVLL signal with magnetic field. The pump intensity was fixed at 0.66~W\,cm$^{-2}$.}
\end{figure}

An example of the sub-Doppler DAVLL signal that we obtained is shown in figure \ref{subdop}. Again, a sharp sub-Doppler feature with a zero-crossing close to the atomic resonance frequency is obtained. However, this  steep sub-Doppler feature is superimposed on a background Doppler-broadened DAVLL signal,  in contrast to the polarization spectroscopy where the Doppler-broadened background absorption is cancelled completely in the difference signal. The variation of the amplitude and gradient with magnetic field and pump beam intensity is shown in figure \ref{BandP}. The amplitude and gradient are comparable to that obtained from polarization spectroscopy with the same pump and probe power (figure \ref{ampgrad}), and follow the same trends. Both the amplitude and gradient depend strongly on the size of the applied magnetic field, increasing to a maximum as the field is increased before decreasing again at very high fields. The optimum gradient should occur when the differential Zeeman shift between the $\sigma^+$ and $\sigma^-$ transitions is equal to the power-broadened linewidth. Here, the maximum gradient is reached at approximately 12\,G, which corresponds to a differential Zeeman shift of 34\,MHz, indicating a small power-broadening of the line at this pump beam intensity. The maximum amplitude occurs at slightly higher magnetic field than the maximum gradient. We have also characterized the variation in the offset between the zero-crossing of the sub-Doppler DAVLL signal and the atomic line centre. This offset depends slightly on the applied magnetic field, varying between 10 and 30~MHz. At a fixed magnetic field, the offset is slightly less stable than for polarization spectroscopy, varying by less than 3~MHz over a similar 1~hour period.

\section{Discussion and Conclusions}

We have applied polarization spectroscopy and sub-Doppler DAVLL to the \probe transition in atomic strontium. Both techniques generate dispersion-shaped lines with sub-Doppler resolution that are suitable for laser locking. With this application in mind, it is instructive to compare the two techniques. For a given laser power, the two techniques produce signals with a similar amplitude and gradient. However, the sub-Doppler DAVLL signal is superimposed on a large Doppler-broadened background, whereas the polarization spectroscopy signal has no Doppler-broadened component. This flat background may contribute to the greater stability in the zero-crossing position that we observed for polarization spectroscopy. Polarization spectroscopy also has the slight advantage that no magnetic field is required. A significant advantage of both techniques is that unlike FM spectroscopy or modulation transfer spectroscopy, neither requires modulation of the laser frequency. In our  laboratory we find that polarization spectroscopy provides a robust and reliable lock for our laser, that is sufficiently stable for us to routinely produce laser cooled strontium atoms. 

  \begin{ack} 
This work was supported by EPSRC grants nos. EP/D070287/1 and EP/D502594/1, and the Royal Society.
 \end{ack}
\bibliographystyle{unsrt.bst}
\bibliography{strontium_pol_spec}
\end{document}